\documentclass[11pt,bookman,fontenc,twoside]{article}
\usepackage{epsfig}
\usepackage{amsmath}
\usepackage{graphicx}
\usepackage{rotating}
\usepackage{hvfloat}
\usepackage{subfigure}
\usepackage[T1]{fontenc}
\usepackage{bookman}
\usepackage{fancyhdr}

\def\singlespace {\smallskipamount=3.75pt plus1pt minus1pt
                  \medskipamount=7.5pt plus2pt minus2pt
                  \bigskipamount=15pt plus4pt minus4pt
                  \normalbaselineskip=15pt plus0pt minus0pt
                  \normallineskip=1pt
                  \normallineskiplimit=0pt
                  \jot=3.75pt
                  {\def\smallskip {\vskip\smallskipamount}}
                  {\def\medskip   {\vskip\medskipamount}}
                  {\def\bigskip   {\vskip\bigskipamount}}
                  {\setbox\strutbox=\hbox{\vrule
                    height10.5pt depth4.5pt width 0pt}}
                  \parskip 7.5pt
                  \normalbaselines}

\def\pr{\prime}
\def\be{\begin{equation}}
\def\lan{\left\langle}
\def\ran{\right\rangle}
\def\ee{\end{equation}}
\def\barr{\begin{array}}
\def\earr{\end{array}}

\def\nn8{\\}
\def\l{\left}
\def\r{\right}
\def\dis{\displaystyle}
\def\ed{\end{document}}

\def\cs{{\bf s}}

\def\spin{\frac{1}{2}}

\def\we{{\widehat {E}}}

\def\wh{{\widehat {H}}}
\def\whh{{\widehat {h}}}
\def\wv{{\widehat {V}}}

\begin{document}

\singlespace
\begin{center}
{\bf Random Interaction Matrix Ensembles in Mesoscopic Physics}
\vskip 0.5cm
Manan Vyas\footnote{{\it E-mail address:} manan@prl.res.in}$^{,}$\footnote{
To appear in the proceedings of the National Seminar on ``New
Frontiers in Nuclear, Hadron and Mesoscopic Physics'', 
eds: V.K.B. Kota, A. Pratap, Allied Publishers, 2010.}
\vskip 0.2cm
Physical Research Laboratory, Ahmedabad 380 009, India
\end{center}
\vskip 1cm
{\footnotesize

\begin{flushleft} 
{\bf Abstract: } 
\end{flushleft}

We analyze several ground state related properties of mesoscopic systems using
the random interaction matrix model EGOE(1+2)-$\cs$ (or RIMM) for many fermion
systems with spin degree of freedom  and the Hamiltonian containing pairing and
exchange interactions in addition to the mean-field one-body and random two-body
parts. RIMM reproduces the essential features of various properties: odd-even
staggering in  ground state energies as a function of particle number, delay in
ground state magnetization and conductance peak spacing distributions. The
analytical formula, we have derived, for the ensemble averaged spectral
variances provides a simple understanding of some of these properties.

\begin{flushleft}
{\bf 1. Introduction to Mesoscopic systems}
\end{flushleft}

Mesoscopic systems are intermediate between microscopic systems (like nuclei and
atoms) and macroscopic bulk matter. Quantum dots and ultrasmall metallic grains
are good examples of mesoscopic systems whose transport properties can be
measured \cite{Im-97,Ja-01}. In these systems, when the electron's phase
coherence length is comparable to or larger than the system size, the system is
called mesoscopic. As the electron phase is preserved in mesoscopic systems,
these are ideal to observe new phenomenon governed by the laws of quantum
mechanics not observed in macroscopic conductors. Also, the transport properties
of mesoscopic systems are readily measured with almost all system parameters
(like the shape and size of the system, number of electrons in the system and
the strength of coupling with the leads) under experimental control. The phase
coherence length increases rapidly with decreasing temperature. For system size
$\sim 100\;\mu$m, the system becomes mesoscopic below $\sim 100$ mK.  

Quantum dots are artificial devices obtained by confining a finite number of
electrons to regions with diameter $\sim 100$ nm by electrostatic potentials.
Typically it consists of $10^9$ real atoms but the number of mobile electrons is
much lower, $\sim 100$. Their level separation is $\sim 10^{-4}$ eV.  If the
transport in the quantum dot is dominated by electron scattering from
impurities, the dot is said to be diffusive and if the transport is dominated by
electron scattering from the structure boundaries, then dot is called ballistic.
The coupling between a dot and its leads is experimentally controllable. When
the dot is strongly coupled to the leads, the electron motion is classical and
the dot is said to be open. In isolated or closed quantum dots, the coupling is
weak and conductance occurs only by tunneling. Also the charge on the closed dot
is quantized and they have discrete excitation spectrum. The tunneling of an
electron into the dot is usually blocked by the classical Coulomb repulsion of
the electrons already in the dot. This phenomenon is called Coulomb blockade.
This repulsion can be overcome by changing the gate voltage. At appropriate gate
voltage, the charge on the dot will fluctuate between $m$ and $m+1$ electrons
giving rise to a peak in the conductance. The oscillations in conductance as a
function of gate voltage are called Coulomb blockade oscillations. At
sufficiently low temperatures, these oscillations turn into sharp peaks. In
Coulomb blockade regime $kT << \Delta << E_c$, the tunneling occurs through a
single resonance in the dot. Here, $T$ is the temperature, $\Delta$ is the mean
single particle level spacing and $E_c$ is the charging energy. Figure
\ref{tunnel} explains the resonant tunneling of a single electron through a
closed quantum dot. 

\hvFloat[floatPos=htb,capWidth=0.5,capPos=r,capVPos=c,objectPos=c]{figure}
{\includegraphics[width=2.5in,height=2.5in]{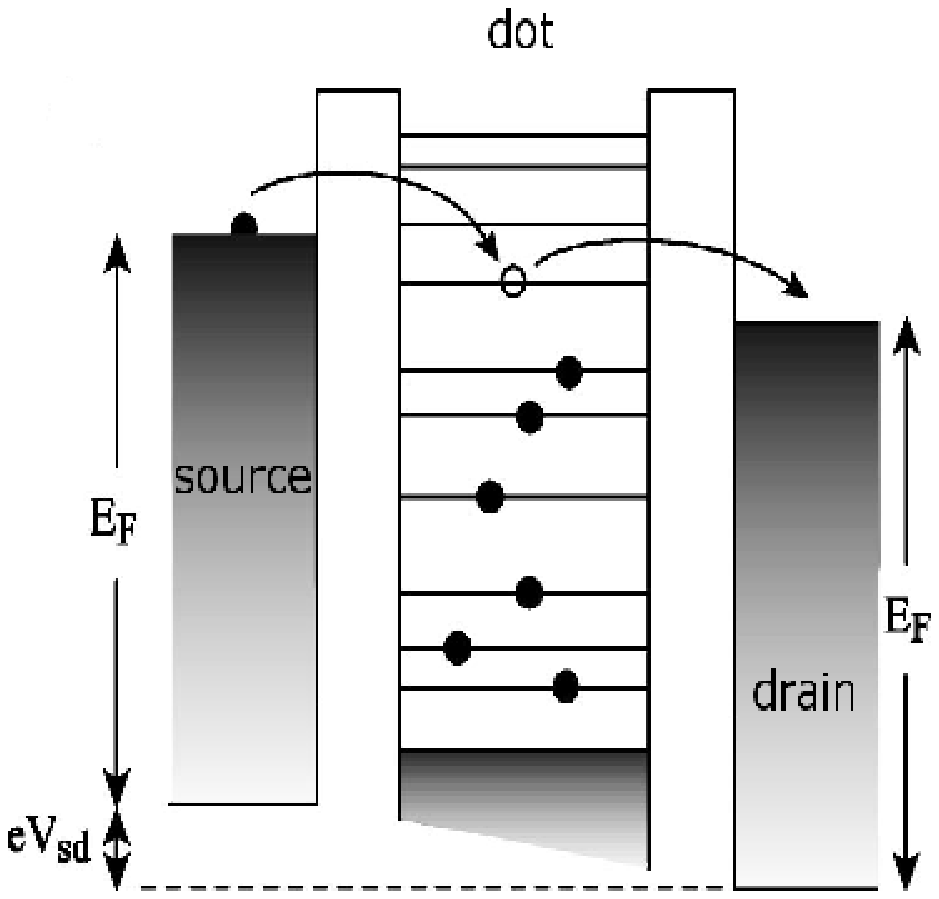}}
{\footnotesize Figure showing conductance of a single electron through an
isolated quantum dot. When Fermi energy $E_F$ of the electron in the source  (s)
and the drain (d) matches the energy of first unoccupied level in the dot,  the
electron tunnels across the barrier into the dots and in response to a  small
source-drain voltage $V_{sd}$, a current will flow.  See text for
details.}{tunnel}

Ultrasmall metallic grains are small pieces of metals of size $\sim 2-10$ nm.
The level separation for nm-size metallic grains is smaller than in quantum dots
of similar size and thus experiments can easily probe the Coulomb blockade
regime in quantum dots. Also, some of the phenomena observed in nm-size metallic
grains are strikingly similar to those seen in quantum dots suggesting that
quantum dots are generic systems for exploring physics of small coherent
structures \cite{Gu-98,Al-00}.

Although the quantum dots contain many electrons, their properties cannot be
obtained by using thermodynamic limit. The description of transport through a
quantum dot at low temperatures in terms of local material constants breaks down
and the whole structure must be treated as a single coherent entity. The quantum
limits of electrical conduction are revealed in quantum dots and conductivity
exhibits statistical properties which reflect the presence of one-body chaos,
quantum interference and electron-electron interaction. The transport properties
of a quantum dot can be measured by coupling it to leads and passing current
through the dot. The conductance through the dots displays mesoscopic
fluctuations as a function of gate voltage, magnetic field and shape
deformation. The techniques used to describe these fluctuations include
semiclassical methods, random matrix theory and supersymmetric methods
\cite{Al-00}. 

Mesoscopic fluctuations are universal dictated only by a few basic symmetries of
the system. It is now widely appreciated that the universal conductance
fluctuations  are intimately related to the universal statistics of finite
isolated  quantum systems whose classical analogs are chaotic
\cite{Ko-01,PW-07}. In describing transport through these coherent systems, we
are interested in quantum manifestations of classical chaos. The link between
classical and quantum chaos was first established in 1984 with
Bohigas-Giannoni-Schmidt conjecture that statistical quantal fluctuations of a
classically chaotic system are described by random matrix theory.

Scattering of electrons from impurities or irregular boundaries leads to single
particle dynamics that are mostly chaotic. Random matrix theory describes the
statistical fluctuations in the universal regime i.e. at energy scales below the
Thouless energy $E=g\Delta$, $g$ is the Thouless conductance. In this universal
regime random matrix theory addresses questions about statistical behavior of
eigenvalues and eigenfunctions rather than their individual description. We
consider a closed mesoscopic system (quantum dot or small metallic grain) with
chaotic single particle dynamics and with large Thouless conductance $g$. Such a
structure is described by an effective Hamiltonian which comprises of a mean
field and two-body interactions preserving spin degree of freedom. For chaotic
isolated mesoscopic systems, randomness of single particle energies leads to
randomness in effective interactions that are two-body in nature. Hence it is
important to invoke the ideas of two-body ensembles to understand and also
predict  properties of these systems theoretically as we shall see ahead.

\begin{flushleft}
{\bf 2. Random matrix theory}
\end{flushleft}

Random matrix theory (RMT) has been established to be fundamental for quantum
systems and beyond; see Fig. \ref{appl}. RMT helps to analyze the statistical
properties of physical systems whose exact Hamiltonian is too complex to be
studied directly. In this paper, we will focus our discussion to applications of
RMT generated by random interactions  to mesoscopic systems. 

\hvFloat[floatPos=htb,capWidth=0.5,capPos=r,capVPos=c,objectPos=c]{figure}
{\includegraphics[width=2in,height=2.5in,angle=-90]{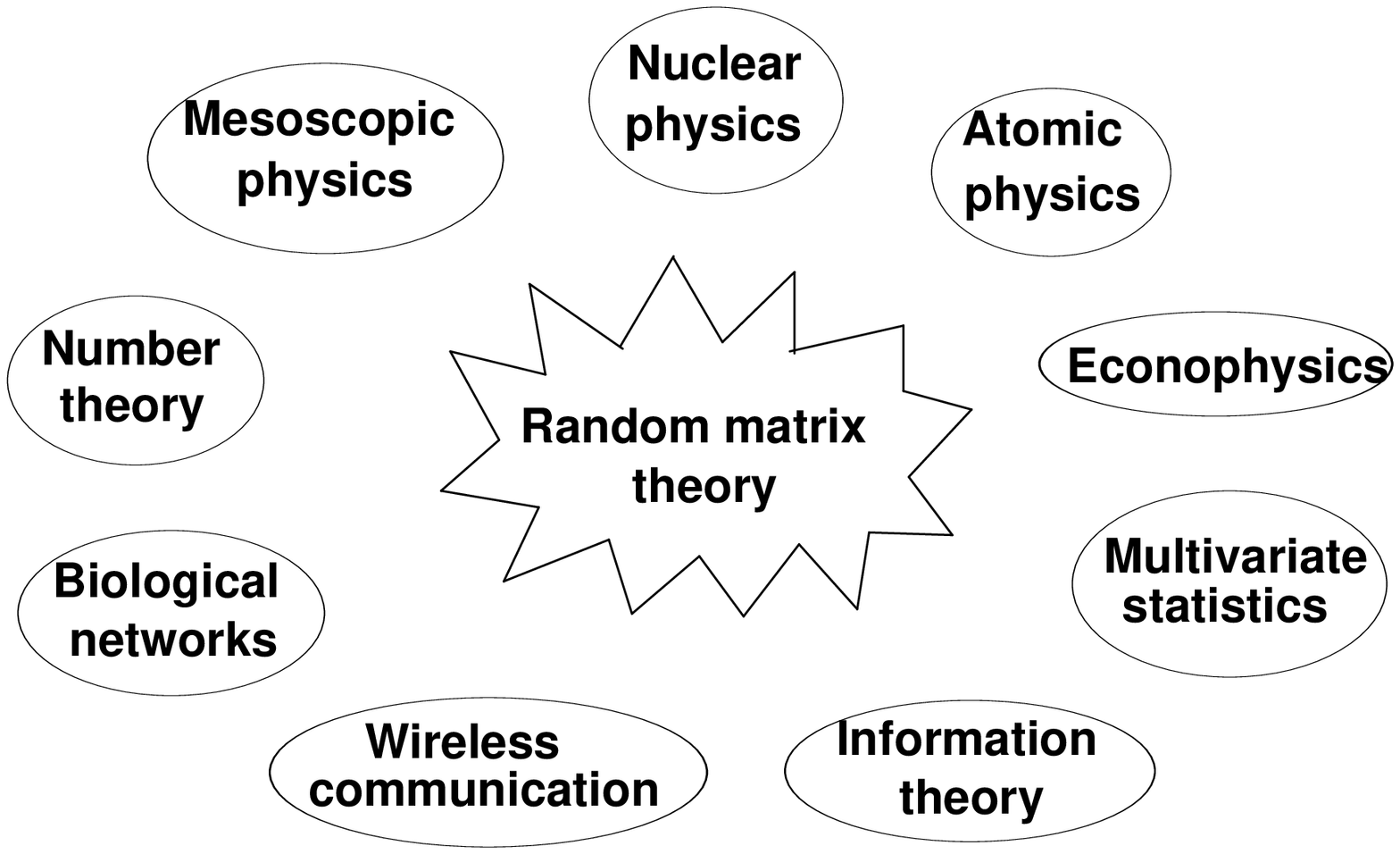}}
{\footnotesize Figure showing wide ranging applications of random matrix 
theory.}{appl}

Classical random matrices belong to three classes: Gaussian orthogonal ensemble
(GOE), Gaussian unitary ensemble (GUE) and Gaussian symplectic ensembles (GSE).
Assuming that the Hamiltonian describing a quantum system is both rotational and
time-reversal invariant, the appropriate ensemble is GOE which is an ensemble of
real-symmetric Hamiltonian matrices. The matrix elements are chosen to be
independent Gaussian random variables with zero center and variance unity
(except that the diagonal matrix elements have variance 2). Then this ensemble
will be invariant under orthogonal transformations and accordingly it is called
GOE. Similarly one can also define GUE and GSE depending upon the global
symmetries of the Hamiltonian of the system. The classical ensembles do not
carry any information beyond the global symmetries (i.e. rotational and
time-reversal invariance) \cite{Po-65,Me-04}.

\hvFloat[floatPos=htb,capWidth=0.5,capPos=r,capVPos=c,objectPos=c]{figure}
{\includegraphics[width=2.25in,height=2.75in]{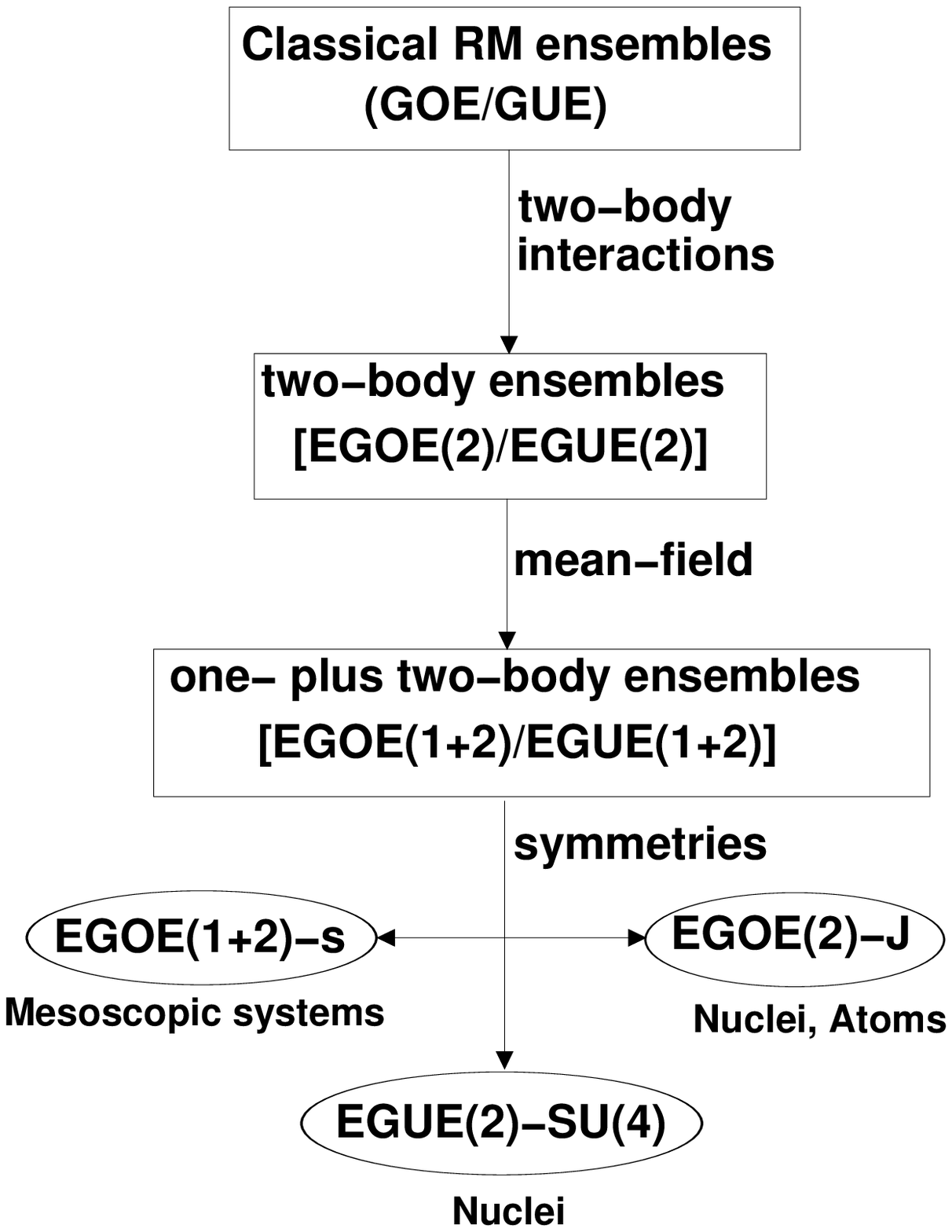}}
{\footnotesize Figure showing the information content of various random matrix 
ensembles. Also shown are the areas in which embedded ensembles with various
symmetries are relevant.}{tbre}

Classical ensembles are generated by $m$-body forces for a $m$-particle system.
However constituents of finite quantum systems interact via two-body
interactions. Therefore, a better random  matrix hypotheses is to consider the
effective interactions to be random. Matrix ensembles generated by random
two-body interactions are called two-body ensembles. As the random matrix
ensemble in two particle spaces is embedded  in the $m$ particle $H$ matrix,
therefore these ensembles are  more generically called embedded ensembles (EE)
\cite{Mo-75}. Note that just as  two-body ensembles it is possible to define
$k$-body ($k < m$) ensembles. Besides two-body interaction, Hamiltonians for
mesoscopic systems  also contain a mean field one-body part and  therefore a
more appropriate  random matrix ensemble is EGOE(1+2), the embedded GOE of one
plus two-body interactions and in more simple terms they are called random
interaction matrix models (RIMM). 

Eigenstates of realistic systems like nuclei, atoms, quantum dots,  small
metallic grains etc. have additional symmetries in addition to the mean field
and two-body interactions. For example, spin ($S$) quantum number is important
for quantum dots and metallic grains. Similarly for atomic nuclei, important are
orbital angular momentum ($L$), spin ($S$) and  isospin ($T$), i.e. $LST$ or
total angular momentum $J$ ($\vec{J}=\vec{L}+\vec{S}$) and isospin, i.e. $JT$.
In some situations just $J$ and $SU(4)$ symmetry are also appropriate. Similarly
for atoms we have $LS$ or $J$ symmetry. Figure \ref{tbre} gives the
hierarchy of the evolution of classical ensembles to EE with symmetries. As
group symmetries define various quantum numbers, in general one has  to consider
EE with group symmetries \cite{Kk-02,Ko-06}.  Numerical as well analytical study
of  these more general ensembles is challenging  due complexities of group
theory, and also in many situations even numerical exploration is quite
complicated. We use a mixture of analytical and numerical methods to make
progress as each method has its own limitations. The first non-trivial and
important (from the point of view of its applications) embedded ensembles are
EE(2)-$\cs$ and EE(1+2)-$\cs$ with spin degree of freedom, for a system of
interacting fermions. In the last decade, the GOE and GUE versions of these
ensemble have received considerable attention \cite{Ko-06,Ko-07,Ma-09,Ma-10}. To
fix the basic ideas involved here let us briefly consider the EGOE(1+2)-$\cs$
ensemble as we will use the results derived for these ensembles to study some
properties of mesoscopic systems. 

\begin{flushleft}
{\bf 3. One- plus two-body random matrix ensembles with spin degree of
freedom}
\end{flushleft}

Let us begin with a system of $m$ ($m > 2$)  fermions  distributed say in
$\Omega$ number of single particle (sp) orbitals each with spin  $\cs=\spin$ so
that the number of sp states $N=2\Omega$; see Fig. \ref{nip} ahead.  For one
plus two-body Hamiltonians preserving $m$ particle spin $S$, the one-body
Hamiltonian is $\whh(1)=\sum_{i=1,2,\ldots,\Omega}\, \epsilon_i n_i$. Here the
orbits $i$ are doubly degenerate, $n_i$ are number operators  and $\epsilon_i$
are sp energies [it is in principle possible to consider $\whh(1)$ with
off-diagonal energies $\epsilon_{ij}$]. Similarly the two-body Hamiltonian
$\wv(2)$ is defined by the two-body matrix elements $V^s_{ijkl}={_a}\lan
(kl)s,m_s \mid \wv(2) \mid (ij)s,m_s\ran_a$ with the two-particle spin $s=0,1$
and they are independent of the $m_s$ quantum number. Note that for $s=1$, only
$i \neq j$ and $k \neq l$ matrix elements exist.  Thus $\wv(2)=\wv^{s=0}(2) +
\wv^{s=1}(2)$ and the $V$ matrix in two particle spaces is a direct sum matrix
with the $s=0$  and $s=1$ space matrices having dimensions $\Omega(\Omega+1)/2$
and $\Omega(\Omega-1)/2$ respectively. Now, EGOE(1+2)-$\cs$ for a given $(m,S)$
system is generated  by defining the two parts of the two-body Hamiltonian to be
independent GOE's [one for $\wv^{s=0}(2)$ and other for $\wv^{s=1}(2)$] in the
2-particle spaces and then propagating the $h(1)+V(2)$ ensemble to the
$m$-particle  spaces with a given spin $S$ by using the geometry (direct product
structure), defined by the embedding algebra of the $m$-particle spaces, 
\be
\{\wh\}_{\mbox{EGOE(1+2)-\cs}} = \whh(1) + \lambda_0\, \{\wv^{s=0}(2)\} +
\lambda_1\, \{\wv^{s=1}(2)\}\;,
\label{eq.def1}
\ee
where $\{\wv--\}$ are GOE and $\lambda_0$ and $\lambda_1$ are the strengths of
the $s=0$ and $s=1$ parts of $\wv(2)$ respectively. Here  $\{\;\}$ denotes
ensemble.

The sp energies $\epsilon_i$ can be fixed \cite{Ko-01} or they can be drawn from
the  eigenvalues of a random  ensemble \cite{St-01} or from the center of a GOE
\cite{Al-00}. Without loss  of generality we put $\Delta=1$ so that $\lambda_0$
and $\lambda_1$ are in the  units of $\Delta$. Thus, EGOE(1+2)-$\cs$  is defined
by the five  parameters $(\Omega, m, S, \lambda_0, \lambda_1)$. The action of
the Hamiltonian operator defined by Eq. (\ref{eq.def1}) on appropriately chosen
fixed-($m,S$) basis states generates the EGOE(1+2)-$\cs$ ensemble in ($m,S$)
spaces. The $H$ matrix dimension $d(m,S)$  for a given $(m,S)$, i.e. number of
levels in the $(m,S)$ space [with each of them being $(2S+1)$-fold degenerate],
is
\be
d(m,S)=\dis\frac{(2S+1)}{(\Omega+1)} { \Omega+1 \choose
m/2+S+1} {\Omega+1 \choose m/2-S}\;,
\label{eq.def2}
\ee
satisfying the sum rule $\sum_S\;(2S+1)\;d(m,S)= {N \choose m}$. For example for
$m=\Omega=8$, the dimensions are 1764, 2352, 720, 63 and 1 for $S=0$, 1, 2, 3
and 4 respectively. Similarly for $m=\Omega=12$, they are 226512, 382239,
196625, 44044, 4214, 143 and 1. Therefore, numerical investigation of these
ensembles beyond $m=\Omega=10$ is not feasible even on the fastest computers
available  due to rapid increase in the dimension of the matrix with increasing
number of particles $m$ and number of sp states $N$. Also the  number of
independent random  variables increase with $N$ and we need to consider
sufficiently large number of members. All these considerations  increase the
computation time manifold. The $H$ matrices can be numerically constructed by
using the formulation for spinless fermions in $M_S$ representation and
projecting the spin using matrices for $S^2$ operator \cite{Ko-06}.
Alternatively it is possible to construct $H$ matrix directly in good $S$ basis
using angular momentum algebra \cite{Tu-06}. Numerical code for constructing the
EGOE(1+2)-$\cs$ ensemble using $M_S$ representation has been developed and
tested \cite{Ko-06}. Using this, several properties of the spin ensemble  have
been investigated in detail, namely pairing correlations \cite{Ma-09} and
transition markers generated by these ensembles \cite{Ma-10}. It is verified in
these studies that properties of EE for spinless fermions extend to ensembles 
with spin \cite{Ko-06,Ma-09,Ma-10}. Now we will briefly discuss some results
obtained for these ensembles.

The fixed-$(m,S)$ level density $\rho^{m,S}(E)$ is numerically established to be
Gaussian \cite{Ko-06,Ma-10,St-01,Ka-00}. Using trace propagation method and
carrying  out ensemble average, exact formula for ensemble averaged spectral
variance i.e. for the variance of ensemble averaged level density generated by
two-body random matrix ensembles which is scalar in the spin space is derived 
\cite{Ma-10} and it is of the form,
\be
\barr{l}
\overline{\sigma^2(m,S)} = \lambda^2 P(\Omega,m,S)\;;\;\;\lambda_0^2=
\lambda_1^2=\lambda^2\;; \\
P(\Omega,m,S)=
\dis\sum_{s=0,1}
\dis\frac{1}{\Omega[\Omega+(-1)^s]/2}
\l[\dis\frac{\Omega+2}{\Omega+1} Q^1(f_s:m,S) +
\dis\frac{\Omega^2+[2+(-1)^s]\Omega+2}
{\Omega^2+[2+(-1)^s]\Omega}\,Q^2(f_s:m,S)\r]\;,
\earr \label{var}
\ee
with $f_0=\{2\}$ and $f_1=\{1^2\}$. Also, 
\be
\barr{rcl}
Q^1(\{2\}:m,S) & = & \l[(\Omega+1) P^0(m,S)/16\r]\,\l[m^x(m+2)/2 + 
\lan S^2\ran\r]\,,\\
Q^2(\{2\}:m,S) & = & \l[\Omega (\Omega+3) P^0(m,S)/32\r]\,\l[m^x(m^x+1) -
\lan S^2\ran\r]\,,\\
Q^1(\{1^2\}:m,S) & = & \dis\frac{(\Omega-1)}{16(\Omega-2)}
\l[(\Omega+2)\,P^1(m,S)
\,P^2(m,S) + 8\Omega (m-1)(\Omega-2m+4) \lan S^2\ran\r]\;, \\
Q^2(\{1^2\}:m,S) & = & \dis\frac{\Omega}{8(\Omega-2)} 
\l.[(3\Omega^2 -7\Omega +6)
(\lan S^2\ran)^2 \r. + 3m(m-2)m^x(m^x-1)(\Omega+1)(\Omega+2)/4 \\
& + &  \l. \lan S^2\ran \l\{-m m^x (5\Omega-3)(\Omega+2)+
\Omega(\Omega-1)(\Omega+1)(\Omega+6)\r\}\r]\;,\\
P^s(m,S) & = & \l[\l\{2-(-1)^s\r\}m \l\{ m+2(-1)^s\r\} - 
(-1)^s 4S(S+1) \r]\;;\;\;s=0,\;1 \\
P^2(m,S) & = & 3 m^x(m-2)/2 - \lan S^2\ran\;,\;\;\;
m^x=\l(\Omega-\dis\frac{m}{2}\r)\;.
\earr \label{var1}
\ee
In Eq. (\ref{var}), the `bar' denotes the ensemble average. We will use the
variance propagator $P$ ahead to explain several properties of mesoscopic
systems. Unlike for a GOE, embedded ensembles generate non-zero cross
correlations between levels with different particle numbers and spins as
randomness is only in the two-particle spaces. The lowest order correlations are
defined by the two-point function,
\be
S(E_i,W_j) = \overline{\rho^{m,S}(E_i)\rho^{m^\pr,S^\pr}(W_j)} - 
\overline{\rho^{m,S}(E_i)}\;\;\overline{\rho^{m^\pr,S^\pr}(W_j)}\;.
\label{eq.twopt}
\ee
For $m \neq m^\pr$ and/or $S \neq S^\pr$, the two-point function generates cross
correlations. The lowest two bivariate moments of the two-point function that
define the lowest order cross correlations are $\Sigma_{11}$ and $\Sigma_{22}$
\cite{Ma-10}. They give cross correlations in energy centroids and spectral
variances respectively. The result in Eq. (\ref{var}) gives exact formula for
normalized cross correlations in energy centroids $\Sigma_{11}$ \cite{Ma-10}. 
Using this, we show in Fig. \ref{corr} results for $\Sigma_{11}$ as a function
of (a) total spin $S$ and (b) particle number $m$. Figure \ref{corr} shows that
the spin dependence of $\Sigma_{11}$ is weak for low $S$ values. Also the
magnitude of $\Sigma_{11}$ increases with increasing $m$. These analytical
results are consistent with the numerical results presented in \cite{Ko-06}
although the magnitudes are slightly different. The exact formula for cross
correlations in variances is more complicated to derive as it involves
evaluation of $\overline{\lan H^2 \ran^{m,S} \lan H^2 \ran^{m^\pr,S^\pr}}$. 
From numerical calculations \cite{Ko-06a}, it is seen that the cross 
correlations in spectral variances are smaller than in centroids. More
significantly, the correlations for EGOE(1+2)-$\cs$ ensemble are found to be
larger compared to that for EE for spinless fermions \cite{Ko-06,Ko-06a} and
therefore we can expect larger correlations for ensembles with $J$ symmetry
compared to those for EGOE(1+2)-$\cs$ \cite{PW-06}.

\hvFloat[floatPos=htb,capWidth=0.5,capPos=r,capVPos=c,objectPos=c]{figure}
{\includegraphics[width=3in,height=2in]{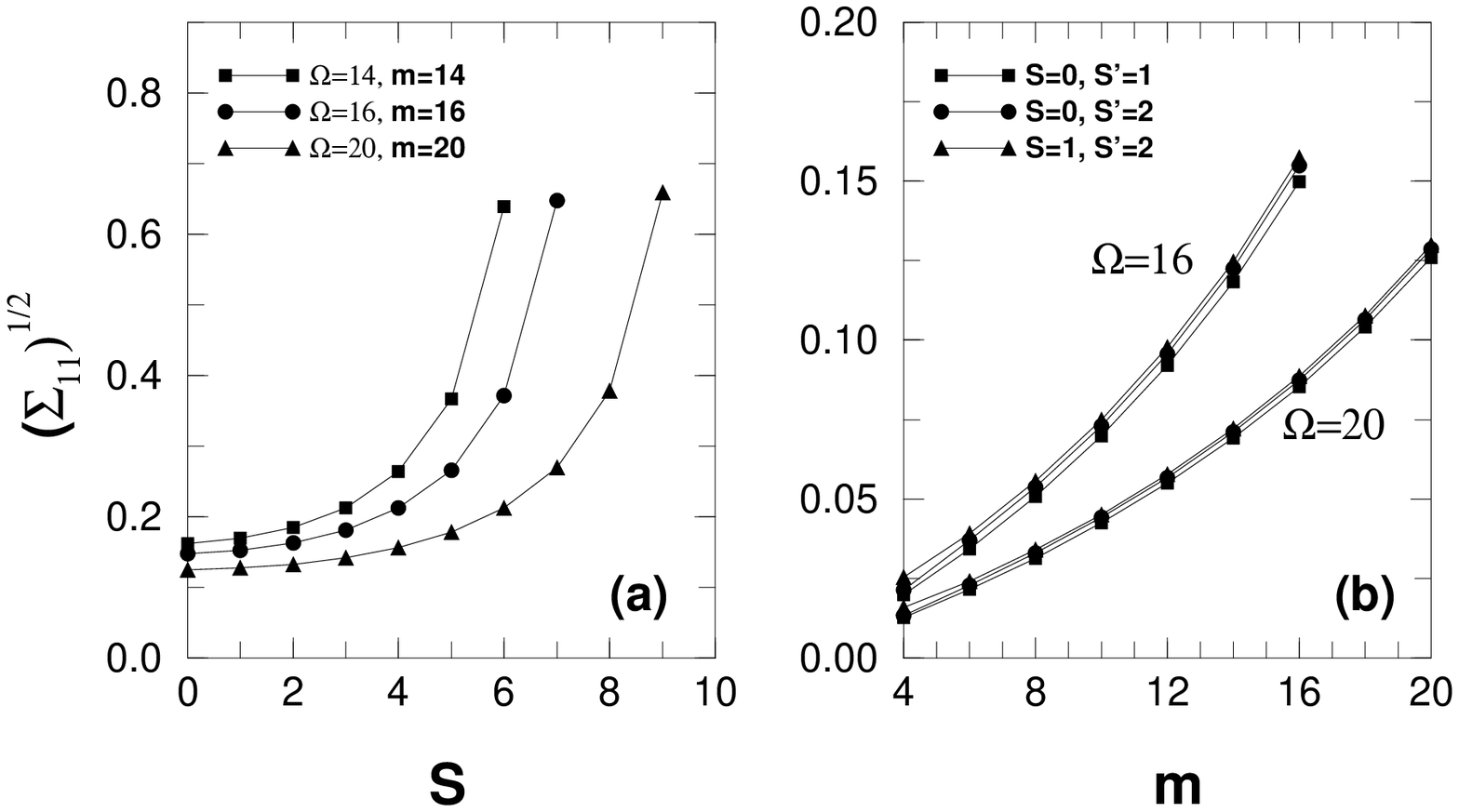}}
{\footnotesize Correlations in energy centroids $\Sigma_{11}$ as a function 
of (a) total spin $S$ and (b) particle number $m$ for EGOE(2)-$\cs$. See text 
for details.}{corr}

In Fig. \ref{dencorr}, we show the correlations between the level densities for
$\Omega=8$ system with $m=m^\pr=8$ and $S=0$, $S^\pr=1$ for a 100 member
EGOE(2)-$\cs$ ensemble. Figure \ref{dencorr} (a) shows the  ensemble average of
the product of two level densities, Fig. \ref{dencorr} (b) shows the product
of ensemble averaged level densities and Fig. \ref{dencorr} (c)  shows the
two-point function $S(E_i,W_j)$. The maximum value of $S(E_i,W_j) \sim 1$\%. We
have verified this result by comparing numerical value of correlations in energy
centroids   with that given by the analytical formula in \cite{Ma-10}. For
ensembles with $JT$-symmetry, the two-point function $S(E_i,W_j) \sim 10$\% for
$^{24}$Mg with $J=2,T=0$ and $J=0,T=0$ \cite{PW-06}. Also the lowest two
bivariate moments  of the two-point function i.e. $\Sigma_{11}$ and
$\Sigma_{22}$ are found to  increase with increasing symmetry when we go from EE
for spinless fermions to EE with spin to EE with SU(4) symmetry \cite{Ma-09a}.
It will be interesting to investigate the increase in fluctuations with
symmetries in detail to understand the role of symmetries in generating chaos.
These non-zero cross correlations should be  subjected to experimental
verification, possibly in some mesoscopic systems. 

\begin{figure}[ht]
    \centering
    \subfigure[$\overline{\rho^{m,S}(E_i)\rho^{m^\pr,S^\pr}(W_j)}$]
    {
        \includegraphics[width=2in,height=2in]{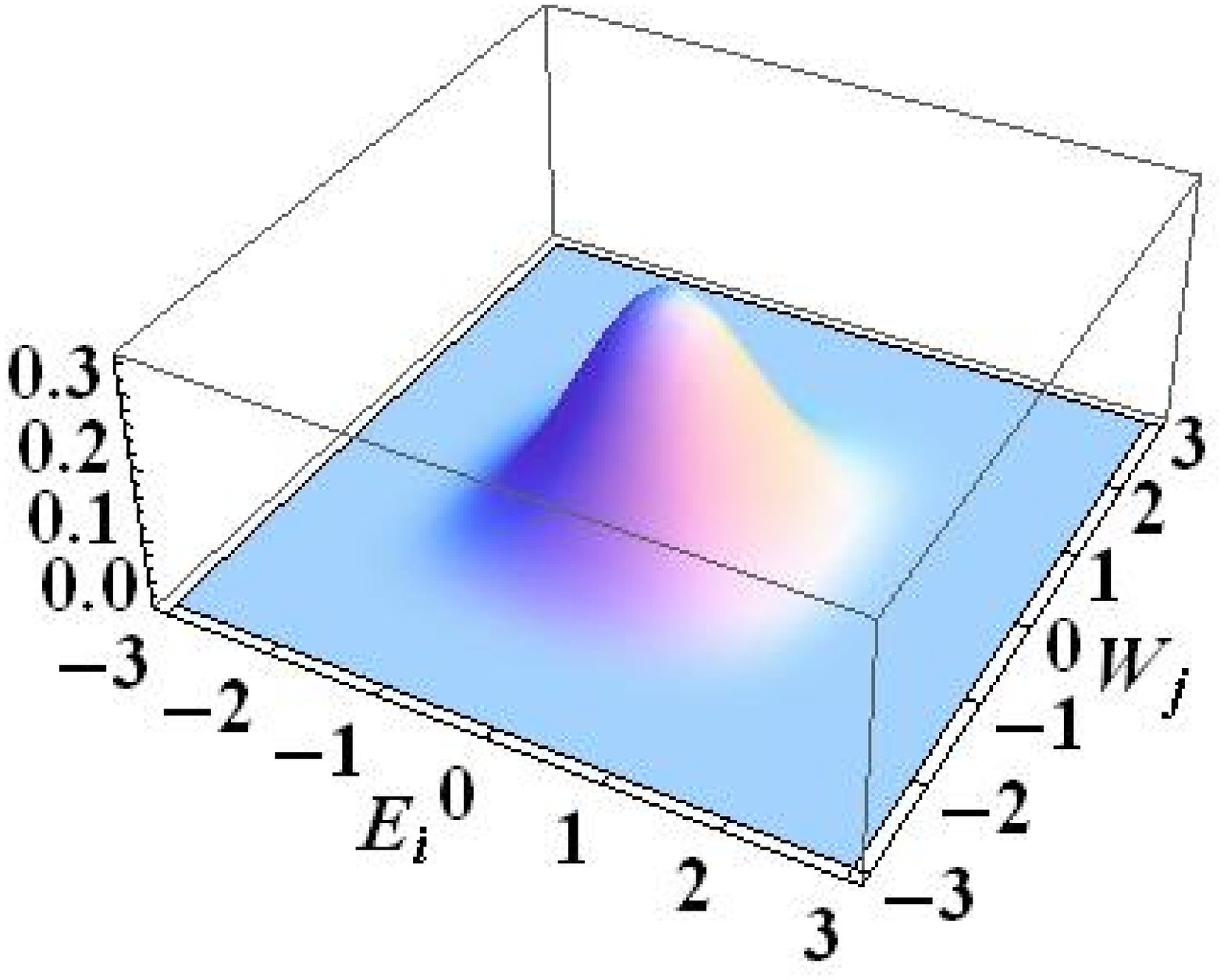}
        \label{ap}
    }
    \subfigure[$\overline{\rho^{m,S}(E_i)}\;\;
    \overline{\rho^{m^\pr,S^\pr}(W_j)}$]
    {
        \includegraphics[width=2in,height=2in]{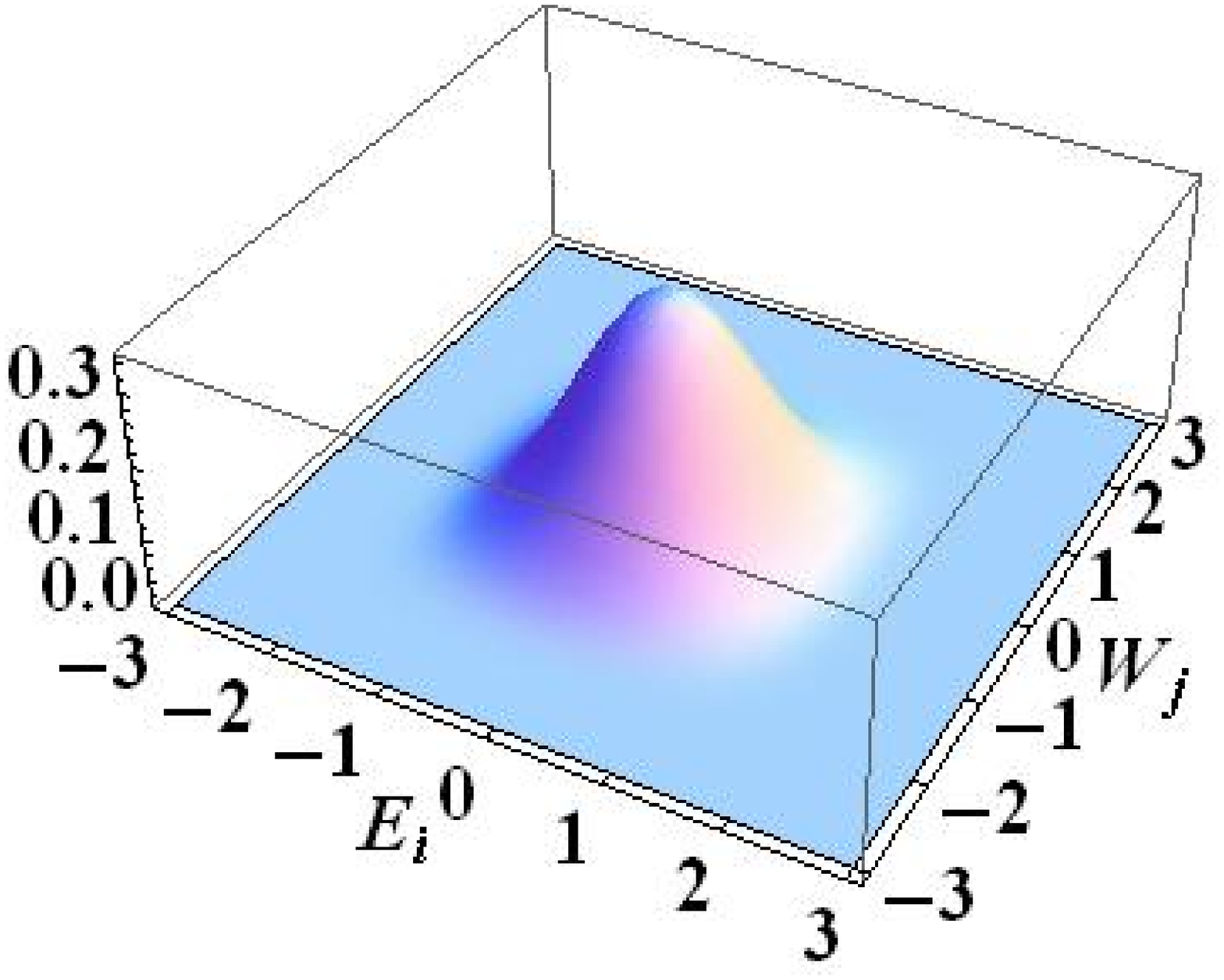}
        \label{pa}
    }\\
    \subfigure[$S(E_i,W_j)$]
    {
        \includegraphics[width=2.25in,height=2in]{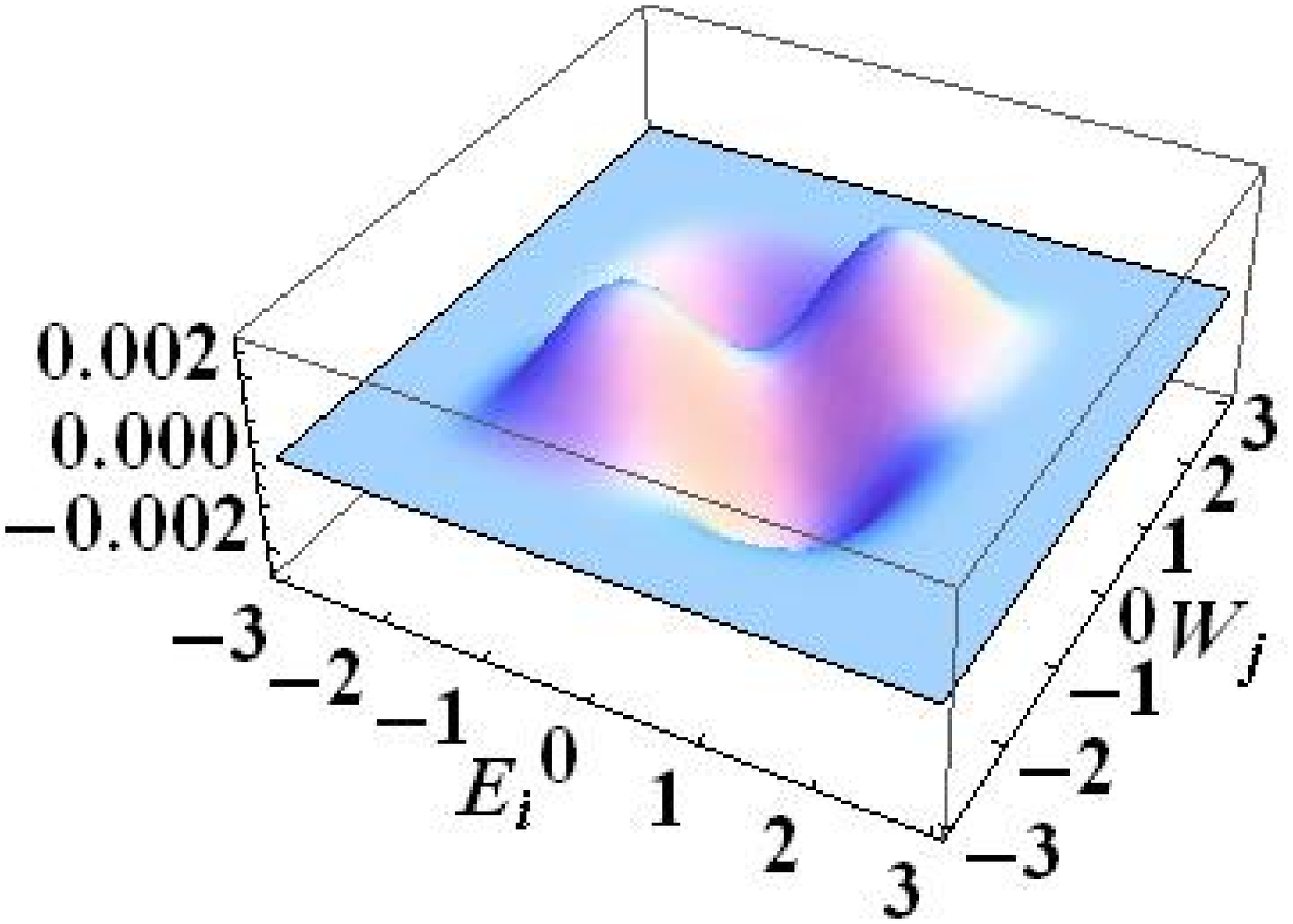}
        \label{dif}
    }
    \caption{\footnotesize Correlations between level densities for $\Omega=m=8$
    system with $S=0$ and $S^\pr=1$ for a 100 member EGOE(2)-$\cs$ ensemble. See
    text for details.}
    \label{dencorr}
\end{figure}

The spin degree of freedom allows us to introduce pairing in EGOE(1+2)-$\cs$ 
\cite{Ma-09}. The pair creation operator $P_i$ for the level $i$ and the 
generalized pair operator are,
\be
P= \dis\frac{1}{\sqrt{2}} \dis\sum_{i=1}^\Omega \l( a^\dagger_{i,\spin}
a^\dagger_{i,\spin} \r)^0 =\dis\sum_{i=1}^\Omega P_i\;.
\label{eq.pair}
\ee  
Therefore in the space defining EGOE(1+2)-$\cs$, the pairing Hamiltonian 
$H_p$ and its matrix elements are,
\be
H_p=PP^\dagger\;,\;\;\;\; _a\lan (k,\ell) s,m_s \mid H_p \mid (i,j) 
s^\pr m_{s^\pr} \ran_a = \delta_{s,0}\delta_{i,j}\delta_{k,\ell}
\delta_{s,s^\pr}
\delta_{m_s,m_{s^\pr}}\;.
\label{eq.pair1}
\ee
The operator $H_p$ defines pairing subspaces denoted by the seniority quantum 
number $v$. Given ($m,S$), spin $S$ is generated by $v$ free particles and 
therefore $v \geq 2S$, $v=m,m-2,m-4,\ldots,2S\;\;\;\;(m \leq \Omega)$.
Now the eigenvalues of $H_p$ are $E_p=(m-v)(2\Omega+2-m-v)/4$. Finally the 
dimension for fixed ($m,v,S$) is $D(m,v,S)=d(m=v,S)-d(m=v-2,S)$.

\hvFloat[floatPos=htb,capWidth=1,capPos=b,capVPos=c,objectPos=c]{figure}
{\includegraphics[width=3.5in,height=2.5in]{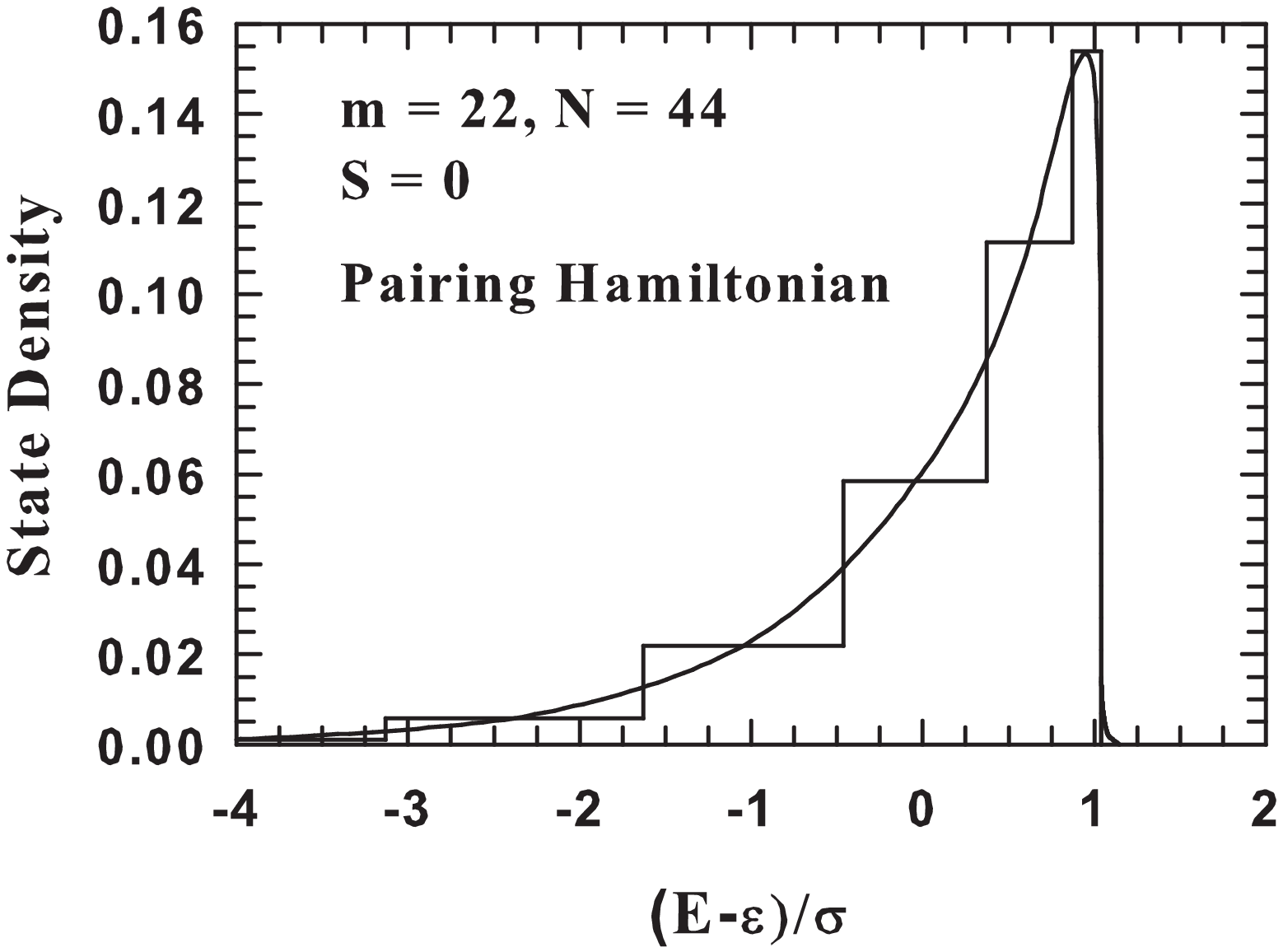}}
{\footnotesize State density for the pairing Hamiltonian $H=-H_p$ for a system
of 22 fermions in $\Omega=22$ orbits ($N=44$) and total spin $S=0$. In the
histogram, $\rho(E)$ for a given $\we=(E-\epsilon)/\sigma$  is plotted with
$\we$ as center with width given by $\Delta\we=\Delta E/\sigma$ (see Eq.
(\ref{eq.dens}) and the following discussion). The smooth curve is obtained by
joining the center points to  guide the eye. A similar plot was shown before by 
Ginocchio \cite{Gi-80} but for a system of identical fermions in a large single
$j$-shell. See text for details.}{hpden}

Before proceeding further we show in Fig. \ref{hpden} a plot of the state 
density generated by the pairing Hamiltonian $H=-H_p$ to demonstrate that it
will be a highly skewed distribution. The dimensions $d(m,S)$ and $D(m,v,S)$
alongwith the energy $E_p$ of $H_p$ will give the normalized density
$\rho(E)$ to be
\be
\rho_{(-H_p)}(E) = \dis\frac{D(m,v,S)}{d(m,S)\;\Delta E}\;;\;\;\;\Delta E =
E_p(m,v+1,S)-E_p(m,v-1,S)=\Omega-v+1\;.
\label{eq.dens}
\ee
Figure \ref{hpden} gives $\rho(\we)$ vs $\we$ plot for $\Omega=22$ (i.e.
$N=44$),  $m=22$ and $S=0$. For this system, the spectrum spread is 132 (note
that  $v_{max}=22$), centroid  $E_c \sim 5.7$ and width $\sigma \sim 6$;  note
that $\we=(E-E_c)/\sigma$. Although the state density for $H_p$ is highly
skewed, the partial densities over the pairing subspaces are Gaussian in the
strong coupling region \cite{Ma-09}. Also the pair expectation values show large
pairing correlations in the ground state (gs) and follow the form derived for
spinless fermions in the strong coupling region \cite{Ma-09}. The transition or
chaos markers generated by EGOE(1+2)-$\cs$ have been studied in detail and their
spin dependence is described using the variance propagator $P$ \cite{Ma-10}. 

A realistic Hamiltonian for mesoscopic systems conserves total spin $S$ and
therefore includes a mean field one-body part, (random) two-body interaction,
pairing $H_p$ and exchange interaction $\hat{S}^2$. In order to  obtain physical
interpretation of the $\hat{S}^2$ operator, we consider the space exchange or
the Majorana operator $M$ that exchanges the spatial coordinates of the
particles and leaves the spin unchanged, i.e.
\be
M \l| i, \alpha; j, \beta \ran = \l| j, \alpha; i, \beta \ran\;.
\label{eq.maj1}
\ee 
In Eq. (\ref{eq.maj1}), labels $i,\;j$ and $\alpha,\;\beta$ respectively denote 
the spatial and spin labels. As the embedding algebra for EGOE(1+2)-$\cs$ is
$U(2\Omega) \supset U(\Omega) \otimes SU(2)$ and $\l| i, \alpha; j, \beta \ran =
(a^\dagger_{i,\alpha}a^\dagger_{j,\beta})\l|0\ran$, we have
\be
2M=C_2\l[U(\Omega)\r] - \Omega \hat{n}\;.
\label{eq.maj2}
\ee
In Eq. (\ref{eq.maj2}), $C_2\l[U(\Omega)\r] = \sum_{i,j,\alpha,\beta}
a^\dagger_{i,\alpha} a_{j,\alpha}  a^\dagger_{j,\beta}a_{i,\beta}$ is the
quadratic Casimir invariant of the  $U(\Omega)$ group,
\be
C_2\l[U(\Omega)\r]=\hat{n}(\Omega+2)-\dis\frac{\hat{n}^2}{2} 
- \hat{S}^2\;.
\label{eq.maj3}
\ee
Combining Eqs. (\ref{eq.maj2}) and (\ref{eq.maj3}), we have finally
\be
M=-\hat{S}^2-\hat{n}\l( \dis\frac{\hat{n}}{4}-1\r)\;.
\label{maj}
\ee
Therefore, the interaction generated by the $\hat{S}^2$ operator is the 
exchange interaction with a number dependent term. This number dependent term 
becomes important when the particle number $m$ changes. The $H$ for isolated 
mesoscopic systems in universal regime has the form (with $\lambda_p$ and 
$\lambda_S$ being positive),
\be
\{\wh(\lambda_0,\lambda_1,\lambda_p,\lambda_S)\} = \whh(1) + 
\lambda_0\, \{\wv^{s=0}(2)\} + \lambda_1\, \{\wv^{s=1}(2)\} -
\lambda_p H_p -\lambda_S \hat{S}^2\;.
\label{ham-mes}
\ee
The constant part arising due to charging energy $E_c$ that depends on the 
number of fermions in the system can be easily incorporated in our model when 
required. For more details on two-body ensembles and mesoscopic systems see
\cite{Gu-98,Al-00,Ko-01,Mi-00}. Now we will turn to some applications of RIMM
defined by Eq. (\ref{ham-mes}) to mesoscopic systems. From now on, 
We drop the `hat' symbol when there is no confusion.

\begin{flushleft}
{\bf 4. Applications of embedded ensembles to mesoscopic systems}
\end{flushleft}

\begin{flushleft}
{\bf 4.1. Odd-even staggering in ground state energies}
\end{flushleft}

For nm-scale Al particles ($5$-$13$ nm in radius), odd-even staggering is
observed in gs energies measured using electron tunneling \cite{Ti-96}. This
phenomenon is normally associated with pairing interaction effects.
Surprisingly, it can also arise from random two-body interaction \cite{Ka-02}.
Odd-even staggering implies that the gs energy of even particle system is larger
than the arithmetic  mean of its odd number members. Then the staggering
indicator $\Delta(m) = [E_{gs}(m+1) + E_{gs}(m-1) - 2E_{gs}(m)]/2$  is a second
derivative of gs energy with particle number $m$. Numerical
calculations by Papenbrock et al \cite{Ka-02} for a 200 member ensemble with 
$\Omega=10$ and $m=3,4,\ldots,17$ have been used to show that random
interaction generate the staggering effect. 
The largest matrix in this calculation has the dimension $d=63504$. It
is important to mention that even with the best available computing facilities,
it is not yet feasible to numerically study the properties of large systems
($\Omega>>10$) modeled by RIMM. 

\hvFloat[floatPos=htb,capWidth=0.5,capPos=r,capVPos=c,objectPos=c]{figure}
{\includegraphics[width=2in,height=2.5in]{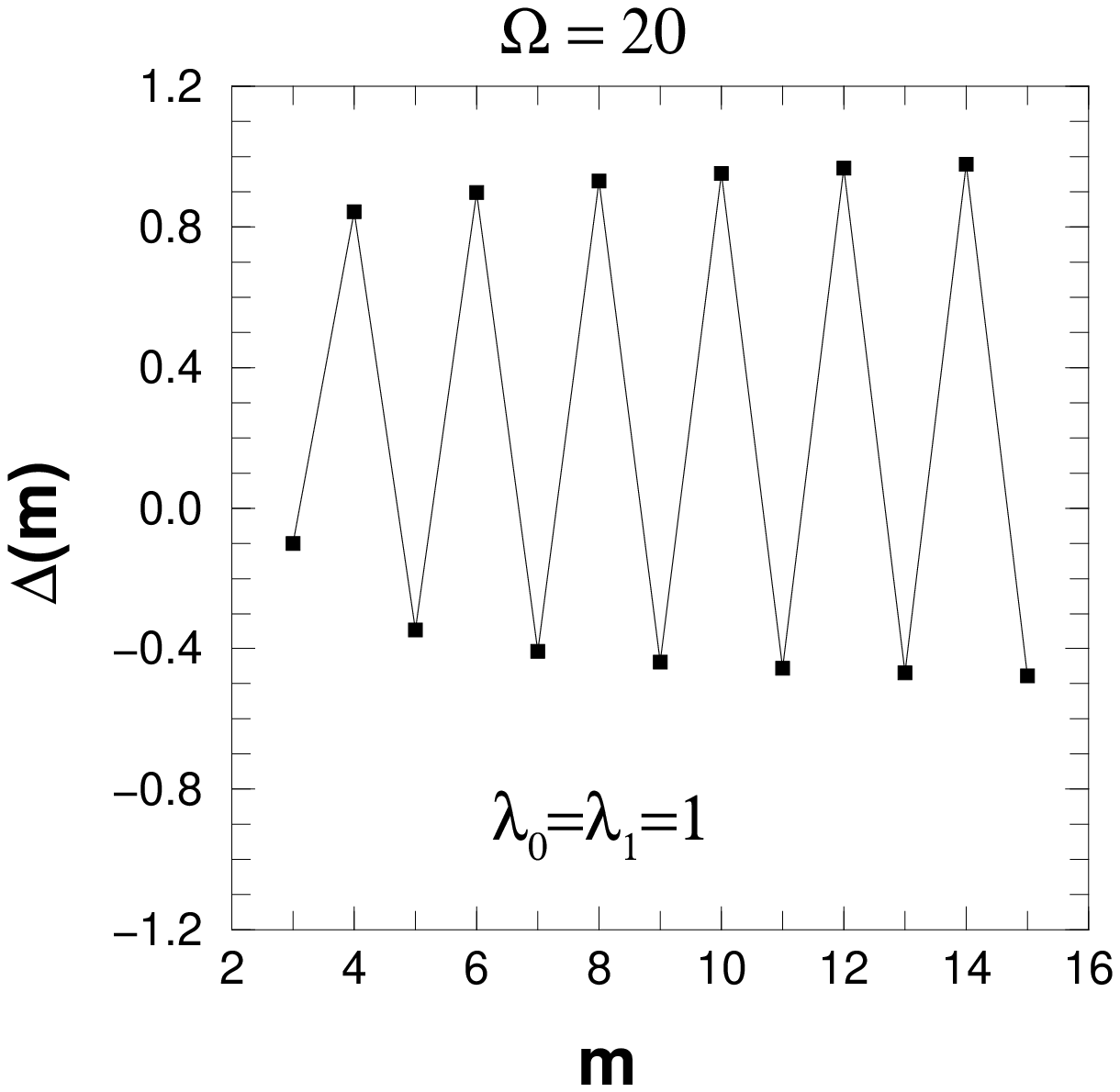}}
{\footnotesize Figure showing staggering in ground state energies with random 
two-body interactions. Calculations are repeated for various combinations of
$\lambda_0$ and $\lambda_1$ values and it is found that the effect is preserved
even when $\lambda_0^2 \neq \lambda_1^2$. See text for details.}{stag}

The eigenvalue density of a system modeled by RIMM (with ensemble averaging) 
is a Gaussian and therefore
the gs energies are largely determined by the widths of the corresponding
Gaussians. As the dependence of gs energies on dimension is logarithmic,
$E_{gs}(m) \propto - \beta\sigma(m,S)$. The prefactor $\beta$ depends on the
details of the deviation of spectral shape from an exact Gaussian form. Though
this is well known in nuclear physics, it was advocated in mesoscopic physics 
in the context of RIMM by
Jacquod and Stone \cite{St-01} and hence we call it JS prescription. Therefore,
with Gaussian fixed-$(m,S)$ densities,  the gs energy is determined by the
spectral widths. Then, using Eq. (\ref{var})  we have
\be
\Delta(m)=\dis\frac{\beta\lambda}{2}[\sqrt{P(\Omega,m+1,S)}+
\sqrt{P(\Omega,m-1,S)}-2\sqrt{P(\Omega,m,S^\pr)}]\;,
\label{gs_eq.1}
\ee 
with $(S,S^\pr)=S_{min}=0$ for even number of particles or $\spin$ for odd 
particle number. We study the staggering phenomenon using our model with
$H(\lambda,\lambda,0,\lambda_S)$ defined by Eq. (\ref{ham-mes}) and using the
analytical formula \cite{Ma-10} for the ensemble averaged variance defined in
Eq. (\ref{var}). Figure \ref{stag} shows the staggering indicator $\Delta(m)$
as a function of particle number $m$ for $\Omega=20$ and $m=2-15$ in units of
$\beta\lambda$. Thus, it is easy to see with the analytical formula for $P$,
that RIMM generates odd-even staggering in gs energies.  

\begin{flushleft}
{\bf 4.2. Delay in ground state magnetization}
\end{flushleft}

We compare the variances for different spins for given ($\Omega,m$) value in
Fig. \ref{gsmag}. It is seen that the variances decrease with increasing spin
independent of the ratio $f=\lambda^2_0/\lambda_1^2$ and therefore the gs energy
(determined by the JS criterion)  will have minimum spin $S$. Thus the exact
formula for the variance propagator \cite{Ma-10} explains  preponderance of gs
with spin $0$ ($m$ even) for mesoscopic systems in a simple way. Also  it is now
well known from a variety of numerical calculations that random  interaction
favor gs spin to be zero (for even $m$) implying that random  interaction
disfavor magnetized ground states. 

\hvFloat[floatPos=htb,capWidth=0.5,capPos=r,capVPos=c,objectPos=c]{figure}
{\includegraphics[width=2in,height=2.5in]{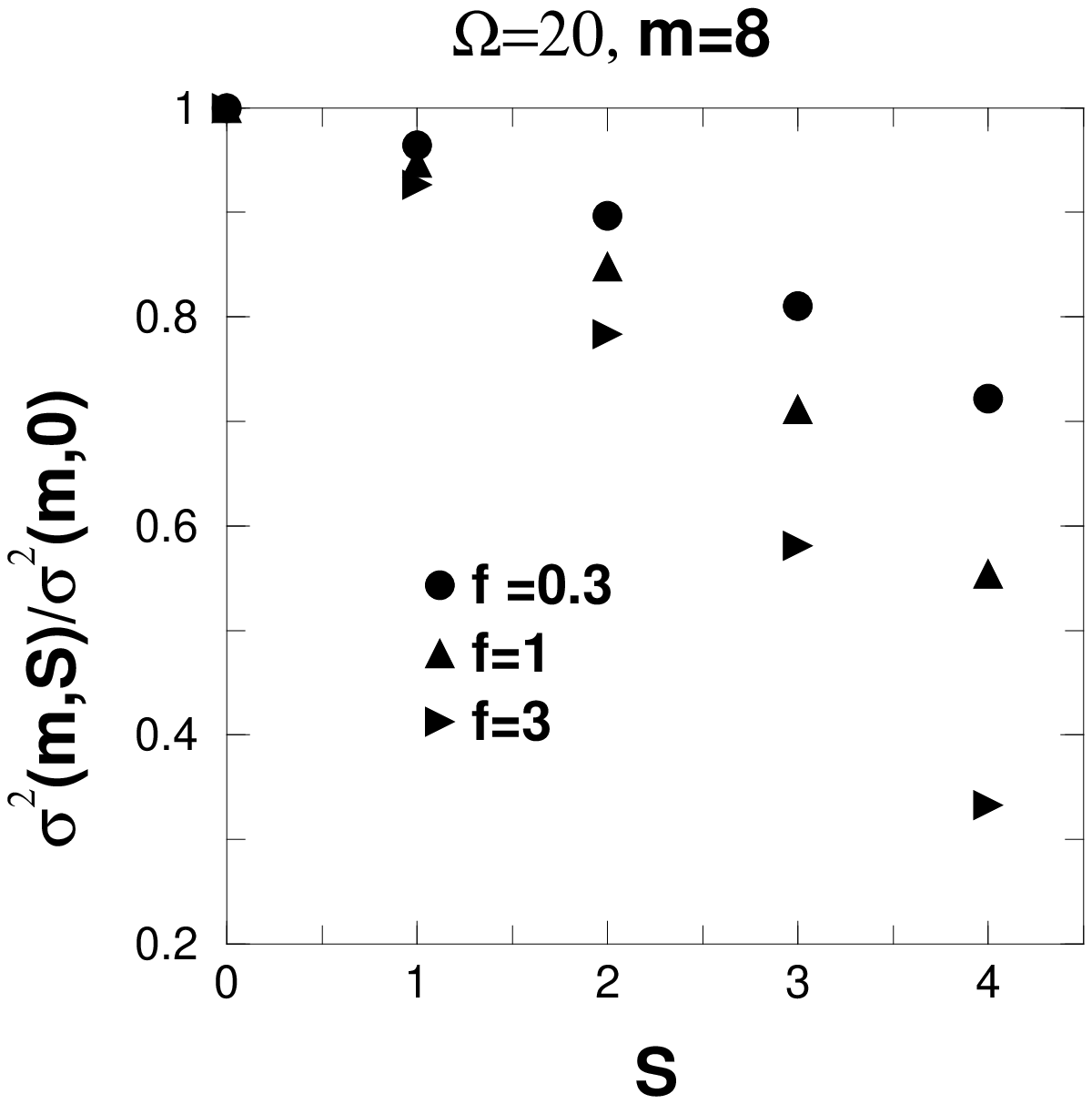}}
{\footnotesize Comparison of spectral variance for various spin sectors for 
$\Omega=20$ and $m=8$ with three different values for  $f = \lambda_0^2 /
\lambda_1^2$ for EGOE(2)-$\cs$. The spectral variances decrease with increasing
spin illustrating that random interaction favor minimal gs  spin.}{gsmag}

The standard Stoner picture of ferromagnetism in itinerant systems is based  on
the competition between one-body kinetic energy [$h(1)$ in Eq. (\ref{ham-mes})]
and the exchange  interaction ($\hat{S}^2$). One-body kinetic energy favors
demagnetized ground states while sufficiently strong repulsive exchange
interaction favors maximum spin to be ground state. In other words, random
interaction disfavor magnetized ground states; see Fig. \ref{gsmag}. As the
minimum spin ground states is favored by random interactions, the Stoner
transition will be delayed in presence of a strong random two-body part in the
Hamiltonian. For a better understanding of these results, we have carried out
numerical calculations for  $\Omega=m=8$ using $H(\lambda,\lambda,0,\lambda_S)$
in Eq. (\ref{ham-mes}). The probability $P(S>0)$ for the gs to be with $S>0$
(for $m$ even) is studied as a function of $\lambda$'s. It is seen from the
results in Fig. \ref{gsspin} that the probability $P(S>0)$ for ground state to
have $S>0$ is very small when $\lambda > \lambda_S$ and it increases with
increasing $\lambda_S$. Figure \ref{gsspin} also gives for a fixed $\lambda$
value, the minimum $\lambda_S$ needed for ground states to have $S>0$ with
$100$\% probability. The results clearly bring out the demagnetizing effect of
random interaction. Similar calculations have been performed in the past for
smaller systems with $\Omega=m=6$ \cite{Ko-06,St-01}. Thus our model explains
the strong bias for low-spin ground states and the delayed ground state
magnetization by random two-body interactions. 

\hvFloat[floatPos=htb,capWidth=0.5,capPos=r,capVPos=c,objectPos=c]{figure}
{\includegraphics[width=3in,height=3in]{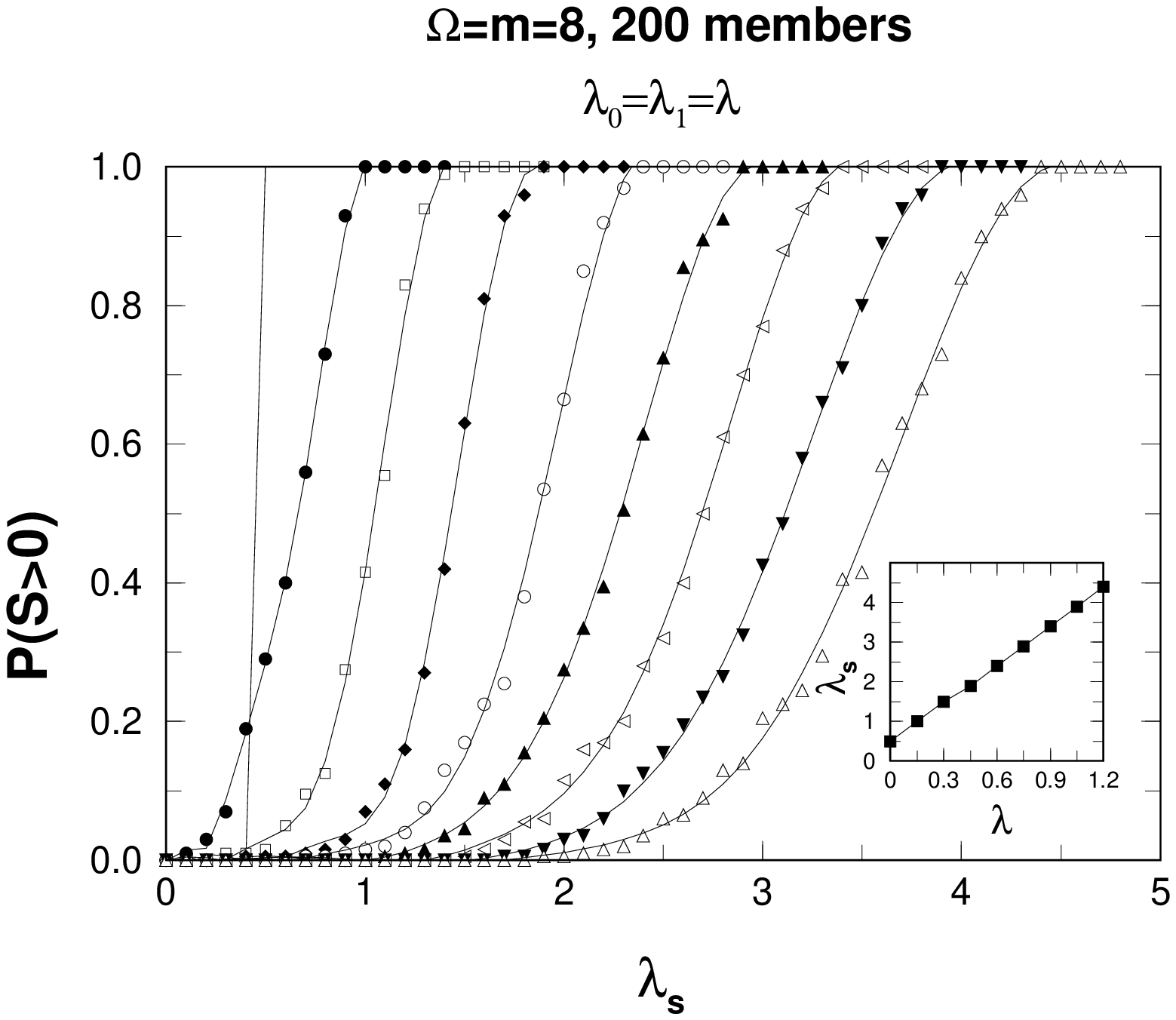}}
{\footnotesize Probability $P(S>0)$ for ground states to have $S>0$ as a 
function of exchange interaction strength $\lambda_S$ for $\lambda=0$ to $1.2$ 
in steps of $0.15$. The calculations are for $200$ member EGOE(2)-$\cs$ ensemble
with $\Omega=m=8$. Inset of figure shows the minimum exchange interaction 
strength $\lambda_S$ required for the ground states to have $S>0$ with $100$\% 
probability as a function of $\lambda$.}{gsspin}

\begin{flushleft}
{\bf 4.3. Conductance peak spacing $(\Delta_2)$ distribution}
\end{flushleft}

Coulomb blockade oscillations yield detailed information about the energy and
wavefunction statistics of mesoscopic systems. We consider a closed mesoscopic
system and study the distribution $P(\Delta_2)$ of spacing $\Delta_2$ between
two neighboring conductance peaks at temperatures less than the average level
spacing. Also our focus is in the strong interaction regime [$\lambda_0 =
\lambda_1 = \lambda \geq 0.3$ in Eq. (\ref{ham-mes})] and we use fixed sp 
energies $\epsilon_i$ \cite{Ma-09}. 

The spacing $\Delta_2$ between the peaks in conductance as a function of the
gate voltage for $T << \Delta$ is second derivative of ground state energies
with respect to the number of particles, 
\be 
\Delta_2 = E_{gs}^{(m+1)} + E_{gs}^{(m-1)} - 2\;E_{gs}^{(m)}\;.
\label{eq.del1}
\ee
In Eq. (\ref{eq.del1}), $E_{gs}^{(m)}$ is ground state energy for a $m$
fermion system. The distribution $P(\Delta_2)$ has been used in the study of
the distribution of conductance peak spacings in chaotic quantum dots 
\cite{Al-05,Al-00} and small metallic grains \cite{Al-08} using chaotic sp 
dynamics. 

Let us first consider non-interacting spinless finite Fermi systems i.e.
$H=h(1)$ with no spin and say  the sp energies are
$\epsilon_i;\;i=1,2,\ldots, N$. Now the ground state energy $E_{gs}^{(m-1)}$
for $m-1$ particles is obtained by filling the sp states from bottom by
applying Pauli principle. Addition of one particle in the system  results in
the gs energy $E_{gs}^{(m)}=E_{gs}^{(m-1)}+\epsilon_m$ and
similarly  $E_{gs}^{(m+1)}=E_{gs}^{(m-1)}+\epsilon_m+\epsilon_{m+1}$, by Pauli
principle. Then using Eq. (\ref{eq.del1}), $\Delta_2=\epsilon_{m+1}
-\epsilon_m$, irrespective of whether $m$ is even or odd. For chaotic systems
it is possible to consider sp energies drawn  from GOE eigenvalues
\cite{Al-00,Al-05}. Therefore $P(\Delta_2)$  corresponds to GOE spacing
distribution $P_W(\Delta)$ - the Wigner distribution. However recent
experiments showed that $P(\Delta_2)$ is a Gaussian in many situations
\cite{Pat-98}. This calls for inclusion of two-body interaction and hence the
importance of RIMM in the study of conductance fluctuations in
mesoscopic systems \cite{Al-00}.

\hvFloat[floatPos=htb,capWidth=0.5,capPos=r,capVPos=c,objectPos=c]{figure}
{\includegraphics[width=3in,height=2.5in]{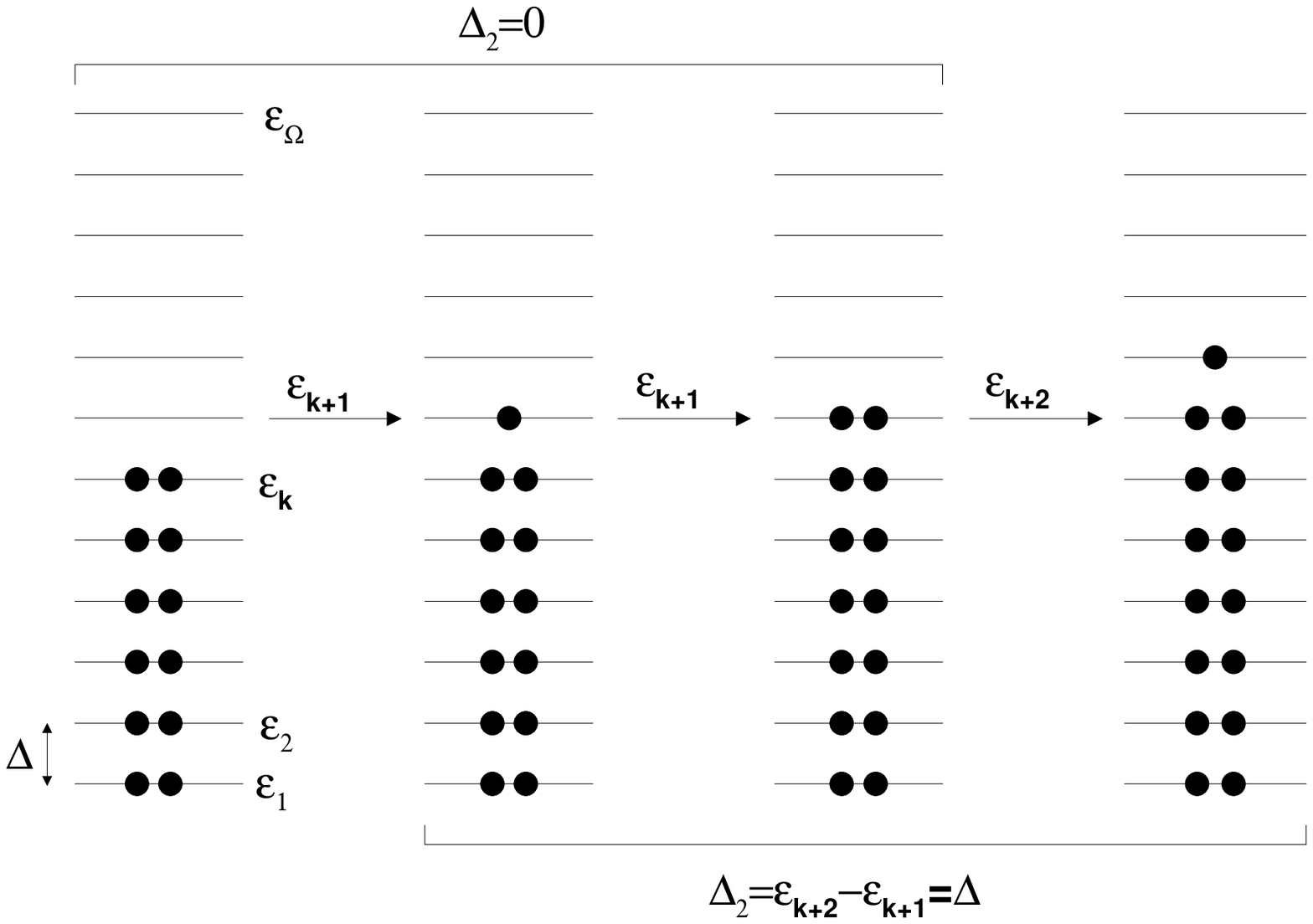}}
{\footnotesize Figure showing $\Delta_2$ values for systems with spin degree of 
freedom. For even-odd-even transitions, $\Delta_2=0$ and for odd-even-odd 
transitions, $\Delta_2=\Delta$. See text for details.}{nip}

As discussed in Section 3, Hamiltonian for interacting electron systems 
conserves total  spin $S$  and thus it is important to consider sp levels that
are doubly degenerate; i.e. spin degree of freedom should be included in $H$.
Again, we start with non-interacting finite  Fermi systems with sp energies
$\epsilon_i$, $i=1,2\ldots,\Omega$ and drawn from a GOE; total number of sp
states $N=2\Omega$. In this scenario $\Delta_2$ depends on whether $m$ is odd
or even. For $m$ odd, say $m=2k+1$, the $(m-1)$ fermion ground state energy
$E_{gs}^{(m-1)}=2\sum_{i=1}^k \epsilon_i$, $E_{gs}^{(m)}=
E_{gs}^{(m-1)}+\epsilon_{k+1}$ and $E_{gs}^{(m+1)}=E_{gs}^{(m-1)}+
2\;\epsilon_{k+1}$ resulting in $\Delta_2=0$. Similar analysis for even $m=2k$
yields $\Delta_2= \epsilon_{k+1}-\epsilon_k$; note that $E_{gs}^{(m)}=
2\sum_{i=1}^k \epsilon_i$, $E_{gs}^{(m-1)}= E_{gs}^{(m)}-\epsilon_{k}$ and
$E_{gs}^{(m+1)}= E_{gs}^{(m)}+\epsilon_{k+1}$. For odd $m$,  $\Delta_2$
corresponds to  even-odd-even transition and $P(\Delta_2)$ is a delta
function. For even $m$, we have odd-even-odd transitions with $P(\Delta_2)$
following Wigner distribution. Figure \ref{nip} gives a pictorial illustration
for $\Delta_2$ calculation for systems with spin. As we need to include, for
real systems, both these transitions, inclusion of spin degree of freedom
gives bimodal distribution for $P(\Delta_2)$,
\be
P(\Delta_2) =\dis\frac{1}{2}\l[\delta(\Delta_2) + P_W(\Delta_2)\r]\;.
\label{eq.del12}
\ee
Convolution of this bimodal form with a Gaussian has been used in the analysis
of  data for quantum dots obtained for situations that  correspond to weak
interactions \cite{Lu-01}. This analysis showed that spin degree of freedom
and pairing correlations are  important for mesoscopic systems. Note that
pairing correlations $(H_p)$ favor minimum spin ground state whereas the
exchange interaction $(-\hat{S}^2)$ tend to maximize the ground state spin. 
Competition between pairing and exchange interaction is equivalent to 
competition between ferromagnetism and superconductivity  \cite{Al-08}. Hence,
it is imperative to study $P(\Delta_2)$ with a  Hamiltonian that includes mean
field one-body part, (random) two-body  interaction, exchange interaction and
pairing (defined by $H_p)$, i.e. $H(\lambda,\lambda,\lambda_p,\lambda_S)$ in Eq.
(\ref{ham-mes}). For small metallic grains, using a microscopic model with
pairing interaction, it was shown in \cite{Al-08} that $P(\Delta_2)$ is bimodal
when pairing interaction is dominant whereas it is unimodal for strong exchange
interaction.

\hvFloat[floatPos=htb,capWidth=1,capPos=r,capVPos=c,objectPos=c]{figure}
{\includegraphics[width=3in,height=2in]{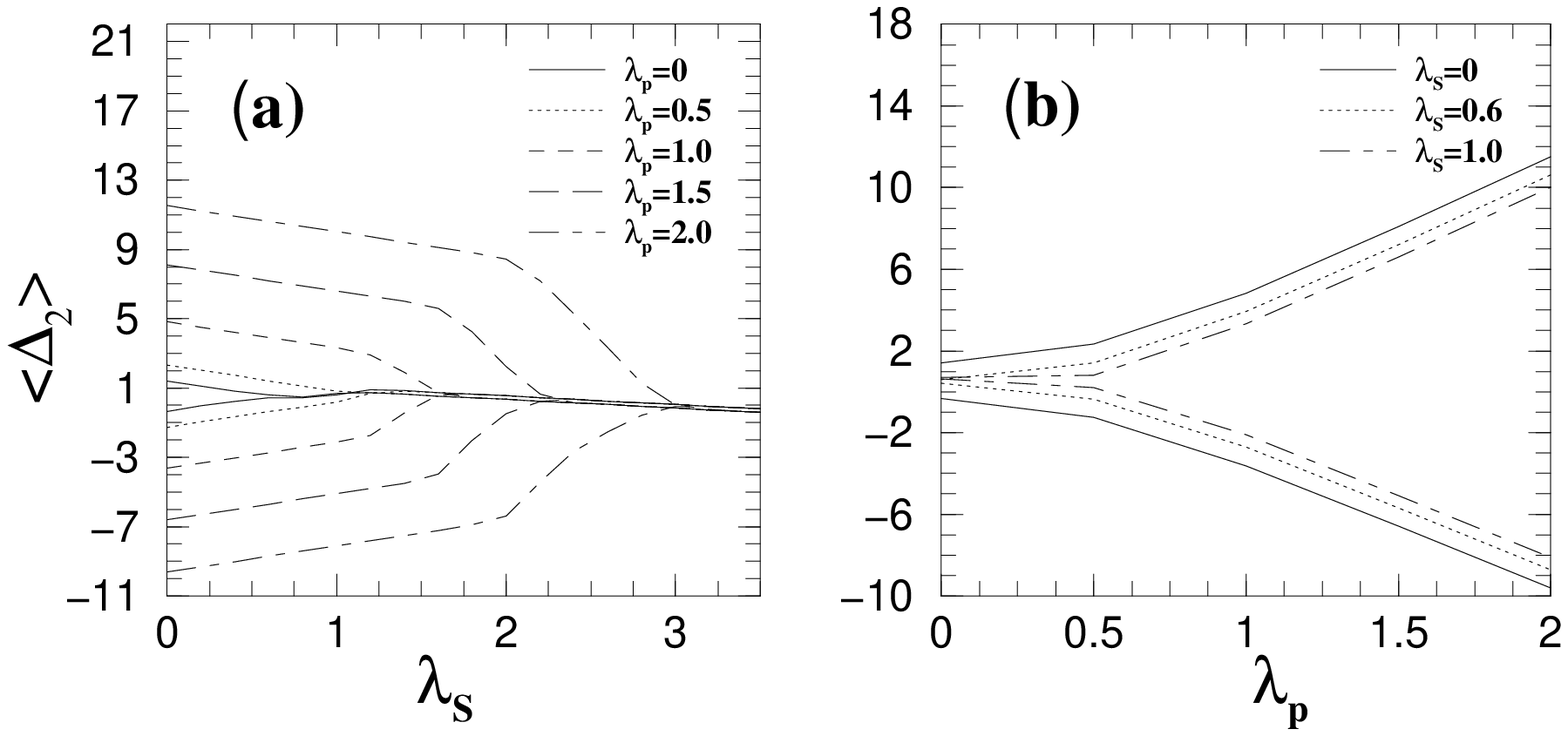}}
{\footnotesize Average peak spacing $\lan \Delta_2 \ran$ (a) as a  function of 
exchange interaction strength $\lambda_S$ for several values of  pairing 
strength $\lambda_p$ and (b) as a function of $\lambda_p$ for several
values of $\lambda_S$, for a 1000 member ensemble with $\Omega=6$. 
The curves in the upper part correspond to $m=4$ ($3 \rightarrow 4  
\rightarrow 5$) and those in the lower part to $m=5$ ($4 \rightarrow 5  
\rightarrow 6$) in Eq. (\ref{eq.del1}). See text for details.}{del2}

Figure \ref{del2}(a) shows the variation of average peak spacing with exchange
interaction strength $\lambda_S$ for several $\lambda_p$ values. The curves in
the upper part correspond to $m=4$ and those in the lower part to $m=5$. As the
exchange strength increases, the average peak spacing  $\lan \Delta_2\ran$ is
almost same for odd-even-odd and even-odd-even  transitions. Value of average
peak spacing and its variation with  $\lambda_S$  is different for odd-even-odd 
and even-odd-even transitions when pairing correlations are strong. The curve
for fixed value of $\lambda_p$ can be divided into two linear regions whose
slopes can be determined considering only exchange interactions, i.e.
$E_{gs}=C_0-\lambda_S\;S\;(S+1)$.  For weak exchange interaction strength,
ground state spin is $0$($1/2$) for $m$ even(odd) and thus for this linear
region, $\lan\Delta_2\ran/\lambda_S \propto -3/2$($3/2$). The linear region
where exchange interactions are dominant,  $\lan\Delta_2\ran/\lambda_S\propto
-1/2$ as ground state spin is $m/2$. Figure \ref{del2}(b) shows the variation of
average peak spacing with pairing strength for several $\lambda_S$ values. It
clearly shows that the separation between the distributions becomes larger with
increasing $\lambda_p$.  These results are in good agreement with the
numerically obtained results for the $P(\Delta_2)$ variation as a function of
$\lambda_p$ and $\lambda_S$ in \cite{Ma-09}. Similar results were reported for
small metallic grains in \cite{Al-08} where a microscopic model is employed
instead of RIMM. Our model with $H$ defined in Eq. (\ref{ham-mes}) thus explains
the interplay between  exchange (favoring ferromagnetism) and pairing (favoring
superconductivity) interaction in the Gaussian domain and can be used for
investigating transport properties of mesoscopic systems.

\begin{flushleft}
{\bf 5. Conclusions}
\end{flushleft}

We have discussed results for mesoscopic systems using EGOE(1+2)-$\cs$ or RIMM
defined by Eq. (\ref{ham-mes}), with pairing and exchange interactions in
addition to mean-field one-body and random two-body parts with spin degree of
freedom. RIMM reproduces the essential features of various ground state related
properties: staggering in ground state energies as a function of particle
number, delay in ground state magnetization and conductance peak spacing
distribution.  The first two properties are studied using analytical methods and
the results are same as those given by large scale numerical calculations. The
RIMM model defined by Eq. (\ref{ham-mes}) can be further generalized by using
random sp energies and more detailed investigation will be carried out in
future. 

\begin{flushleft}
{\bf Acknowledgements}
\end{flushleft}

Thanks are due to V.K.B. Kota for reading a draft of the article and for many 
fruitful discussions. Thanks are also due to N.D. Chavda for useful discussions.
Results in Figs. \ref{dencorr} and \ref{gsspin} are generated using PRL's 20
node cluster computer and thanks are due to computer center staff for their
help.

}
\ed